\documentclass[%
aps,
 pra,%
 amsmath,amssymb,
 reprint,%
 superscriptaddress,
]{revtex4-1}

\usepackage{dcolumn}
\usepackage{graphicx}
\usepackage{lineno}
\usepackage{bm}%
\usepackage[pdftex,colorlinks=true,bookmarks=false,citecolor=blue,urlcolor=blue]{hyperref}
\usepackage{upgreek}

\usepackage{titlesec} 

\begin{document}

\title{\vspace{-15mm}\fontsize{19pt}{10pt}\selectfont\textbf{Semi-nonlinear nanophotonic waveguides for\\
		highly efficient second-harmonic generation}} 

\author{Rui Luo}
\affiliation{Institute of Optics, University of Rochester, Rochester, NY 14627}

\author{Yang He}
\affiliation{Department of Electrical and Computer Engineering, University of Rochester, Rochester, NY 14627}

\author{Hanxiao Liang}
\affiliation{Department of Electrical and Computer Engineering, University of Rochester, Rochester, NY 14627}

\author{Mingxiao Li}
\affiliation{Department of Electrical and Computer Engineering, University of Rochester, Rochester, NY 14627}

\author{Qiang Lin}
\email[Electronic mail: ]{qiang.lin@rochester.edu}
\affiliation{Institute of Optics, University of Rochester, Rochester, NY 14627}
\affiliation{Department of Electrical and Computer Engineering, University of Rochester, Rochester, NY 14627}

\date{\today}



\begin{abstract}
	
Quadratic optical parametric processes form the foundation for a variety of applications related to classical and quantum frequency conversion, which have attracted significant interest recently in on-chip implementation. These processes rely on phase matching among the interacting guided modes, and refractive index engineering has been a primary approach for this purpose. Unfortunately, the modal phase matching approaches developed so far only produce parametric generation with fairly low efficiencies, due to the intrinsic modal mismatch of spatial symmetries. Here we propose a universal design and operation principle for highly efficient optical parametric generation on integrated photonic platforms. By introducing spatial symmetry breaking into the optical nonlinearity of the device, we are able to dramatically enhance the nonlinear parametric interaction to realize an extremely high efficiency. We employ this approach to design and fabricate a heterogeneous titanium oxide/lithium niobate nanophotonic waveguide that is able to offer second-harmonic generation with a theoretical conversion efficiency as high as 2900$\%~\rm {\rm {W^{-1}} cm^{-2}}$, which enables us to experimentally achieve a conversion efficiency of 36.0\%~$\rm {W^{-1}}$ in a waveguide only 2.35 mm long, corresponding to a recorded normalized efficiency of 650$\%~\rm {W^{-1}}cm^{-2}$ that is significantly beyond the reach of conventional modal phase matching approaches. Unlike nonlinearity domain engineering that is material selective, the proposed operation principle can be flexibly applied to any other on-chip quadratic nonlinear platform to support ultra-highly efficient optical parametric interactions, thus opening up a great avenue towards extreme nonlinear and quantum optics with great potentials for broad applications in energy efficient nonlinear and quantum photonic signal processing.

\end{abstract}

\maketitle 


Quadratic optical parametric processes via a $\chi^{(2)}$ nonlinearity have attracted long-lasting interest ever since the first observation of second-harmonic generation (SHG) \cite{BoydBook, Bloembergen00}. Their intriguing capability of creating new light through elastic photon-photon scattering forms a crucial foundation for a wide variety of applications ranging from photonic signal processing \cite{Fejer06, Willner14}, tunable coherent radiation \cite{Dunn99}, frequency metrology \cite{Cundiff03}, optical microscopy \cite{Dong11}, to quantum information processing \cite{Pan12}. Recently, tremendous interest has been attracted to developing optical parametric processes on various chip-scale platforms \cite{Lipson11, Bres17, Pavesi12, Noda14, Barclay16, Vuckovic09, Vuckovic14, Leo11, Tang11, Solomon14, Tang16Optica, Bowers18, Pertsch15, Cheng16, Loncar17OE, Luo17OE, Buse17, Luo18arXiv, Xiao18, Bowers16, Fathpour16PPLN, Loncar18CLEO}, which show great promise to significantly enhance the nonlinear effects by tight confinement of optical fields.

The efficiency of an optical parametric process relies essentially on phase matching among the interacting waves to sustain coherent nonlinear interaction. To date, phase matching is dominantly realized via either domain engineering of nonlinear susceptibility \cite{Helmy11, Fejer07, Shur15} or refractive-index engineering of interacting optical modes \cite{Helmy11}. Domain engineering is an efficient approach to achieve quasi-phase matching, which, however, is fairly material limited. For example, periodic poling \cite{Byer92, Watanabe93, Fejer95, Kato97, Fejer02typical, Fejer02higheff, Fejer07, Fejer10, Shur15} and orientation patterned growth \cite{Helmy11, Yoo95, Fejer01, Tassev13} are currently implemented only to ferroelectrics and III-V semiconductors, respectively. Domain engineering also has stringent requirement on the domain uniformity \cite{Fejer07, Shur15, Byer92}, which imposes a challenge for applications that require small domain periods \cite{Shur15, Fathpour16PPLN, Bowers16, Loncar18CLEO} or clean parametric noise background \cite{Fejer10}. On the other hand, modal index engineering is able to achieve exact phase matching \cite{Helmy11}, which can be flexibly implemented particularly to an on-chip platform that supports rich guided mode species \cite{Vuckovic09, Noda14, Lipson11, Leo11, Tang11, Vuckovic14, Solomon14, Barclay16, Tang16Optica, Bres17, Pertsch15, Cheng16, Loncar17OE, Luo17OE, Buse17, Luo18arXiv, Xiao18, Bowers18}. Unfortunately, modal phase matching suffers from significant mode field mismatch among the interacting optical modes, which seriously degrades the nonlinear conversion efficiency \cite{Vuckovic09, Lipson11, Leo11, Tang11, Noda14, Vuckovic14, Solomon14, Barclay16, Tang16Optica, Bres17, Pertsch15, Cheng16, Loncar17OE, Luo17OE, Buse17, Luo18arXiv, Xiao18, Bowers18}.

\begin{figure*}[t!]
	\centering\includegraphics[width=2\columnwidth]{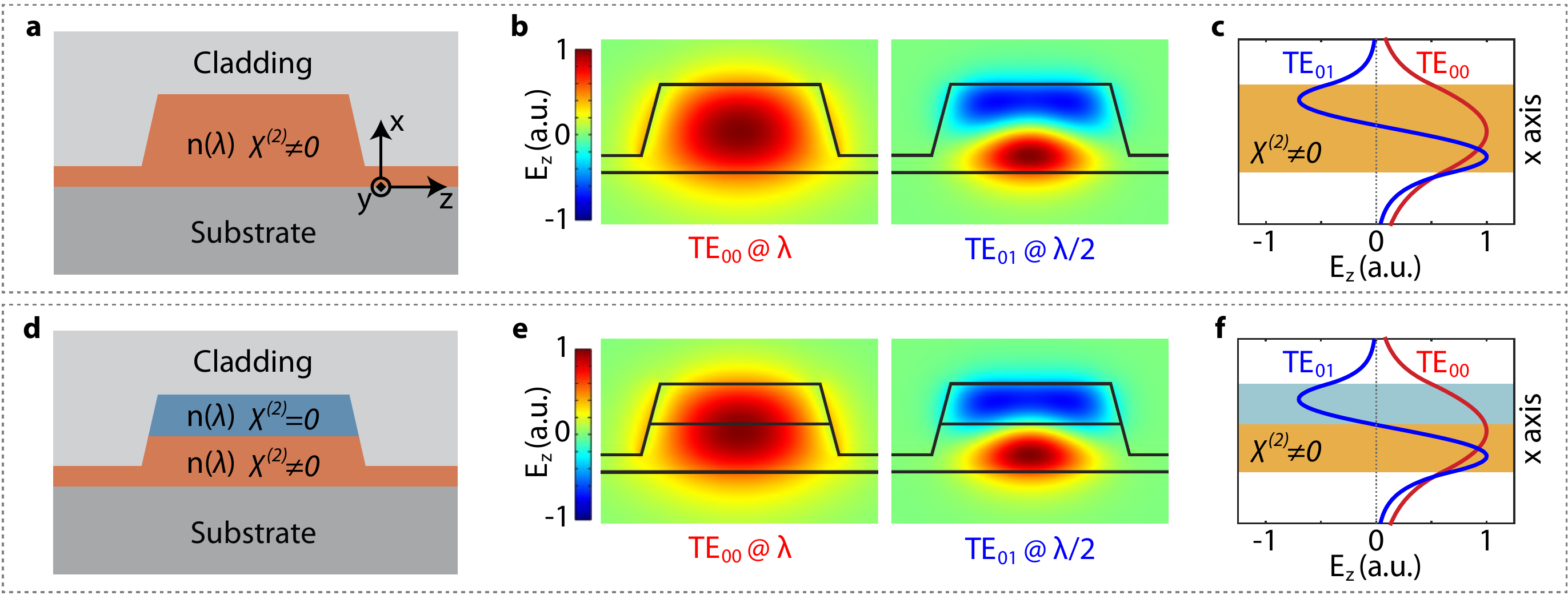}
	\caption{Illustration of the design principle of semi-nonlinear waveguides for highly efficient SHG. {\bf a,} Schematic of a monolithic nanophotonic waveguide with a $\chi^{(2)}$ nonlinearity. {\bf b,} Optical field ($E_z$) profiles of two phase-matched modes, TE$_{00}$ at the fundamental wavelength $\lambda$ and TE$_{01}$ at the half wavelength $\lambda/2$. {\bf c,} $E_z$ as a function of the vertical position x, at the center of the waveguide, with the orange shaded area indicating the $\chi^{(2)}$ material. {\bf d,} Schematic of a semi-nonlinear waveguide composed of two core materials, with both sharing the same linear refractive index $n(\lambda)$, while only the bottom layer has a non-vanished $\chi^{(2)}$ nonlinearity. {\bf e,f,} counterparts of {\bf b} and {\bf c}, respectively, for the semi-nonlinear waveguide. } \label{Fig1}
\end{figure*}
Here we propose a universal design and operation principle of semi-nonlinear waveguides for highly efficient optical parametric generation on integrated photonic platforms. The proposed heterogeneous nanophotonic waveguides consist of a nonlinear medium exciting parametric processes and a linear medium assisting exact phase matching, which are able to combine elegantly a large nonlinearity, a strong optical confinement, and a good spatial mode match together in a single device, resulting in dramatically enhanced nonlinear parametric generation with an extremely high efficiency. To demonstrate our proposed principle, we employ a nanophotonic waveguide composed of a highly nonlinear $\chi^{(2)}$ material lithium niobate (LN) and a linear optical material amorphous titanium oxide ($\rm {TiO_2}$). The dramatically enhanced nonlinear optical interaction in the device leads to a theoretical normalized conversion efficiency as high as 2900$\%~\rm {\rm {W^{-1}} cm^{-2}}$ for SHG, which enables us to experimentally achieve a conversion efficiency of 36.0\%~$\rm {W^{-1}}$ in a waveguide only 2.35 mm long, corresponding to an experimentally recorded normalized conversion efficiency of 650$\%~\rm {W^{-1}}cm^{-2}$. The proposed operation principle can be flexibly applied to any other on-chip quadratic nonlinear platform to support ultra-highly efficient optical parametric interactions, thus opening up a great avenue towards extreme nonlinear and quantum optics with great potentials for broad applications in energy efficient nonlinear and quantum photonic signal processing.

\section{Concept illustration}

In a quadratic nonlinear waveguide, it is well-known that the phase velocities can be exactly matched among different guided modes to support coherent nonlinear interaction \cite{Lipson11, Leo11, Tang11, Solomon14, Barclay16, Tang16Optica, Bres17, Pertsch15, Cheng16, Tang16Optica, Loncar17OE, Luo17OE, Buse17, Luo18arXiv, Xiao18, Bowers18}. When the phase matching condition is satisfied, the efficiency of SHG is given by the following expression (for a lossless waveguide without pump depletion) \cite{BoydBook}
\begin{equation}
\eta \equiv \frac{P_2}{P_1^2 L^2} =\frac{8\pi^2}{\epsilon_0 c n_1^2 n_2 \lambda^2} \frac{\zeta^2 d_{\rm eff}^2}{A_{\rm eff}}, \label{eta}
\end{equation}
where $P_1$ and $P_2$ are the powers of the input fundamental-frequency (FF) mode and the produced second-harmonic (SH) mode, respectively. $L$ is the waveguide length, $d_{\rm eff}$ represents the effective nonlinear susceptibility, $\lambda$ is the fundamental pump wavelength, and $n_1$ and $n_2$ are the effective refractive indices of the FF and SH modes, respectively. $\epsilon_0$ and $c$ are the permittivity and light speed, respectively, in vacuum. In Eq.~(\ref{eta}), $A_{\rm eff} \equiv (A_{1}^2 A_{2})^{\frac{1}{3}}$ is the effective mode area where $A_{i} = \frac{ (\int_{\rm all} |\vec{E}_i|^2 dxdz)^3 }{|\int_{\chi^{(2)}} |\vec{E}_i|^2 \vec{E}_i dxdz|^2}$,~($i=1,2$), and $\zeta$ represents the spatial mode overlap factor between the FF and SH modes, given as
\begin{equation}
\zeta = \frac{ \int_{\chi^{(2)}} (E_{1z}^*)^2 E_{2z} dxdz}{|\int_{\chi^{(2)}}|\vec{E}_1|^2 \vec{E}_1 dxdz|^{\frac{2}{3}}   |\int_{\chi^{(2)}} |\vec{E}_2|^2 \vec{E}_2 dxdz|^{\frac{1}{3}}}, \label{zeta}
\end{equation}
where $\int_{\chi^{(2)}}$ and $\int_{all}$ denote two-dimensional integrations over the $\chi^{(2)}$ material and all space, respectively. $\vec{E}_1(x,z)$ and $\vec{E}_2(x,z)$ are the electric fields of the FF and SH modes, respectively, and $E_{1z}$ and $E_{2z}$ are their z-components. Here we have assumed a type-0 process where the FF and SH modes are both quasi-transverse-electric (quasi-TE) modes with electric fields dominantly lying in the device plane (for example, see Fig.~\ref{Fig1}a,b).

\begin{figure*}[t!]
	\centering\includegraphics[width=2\columnwidth]{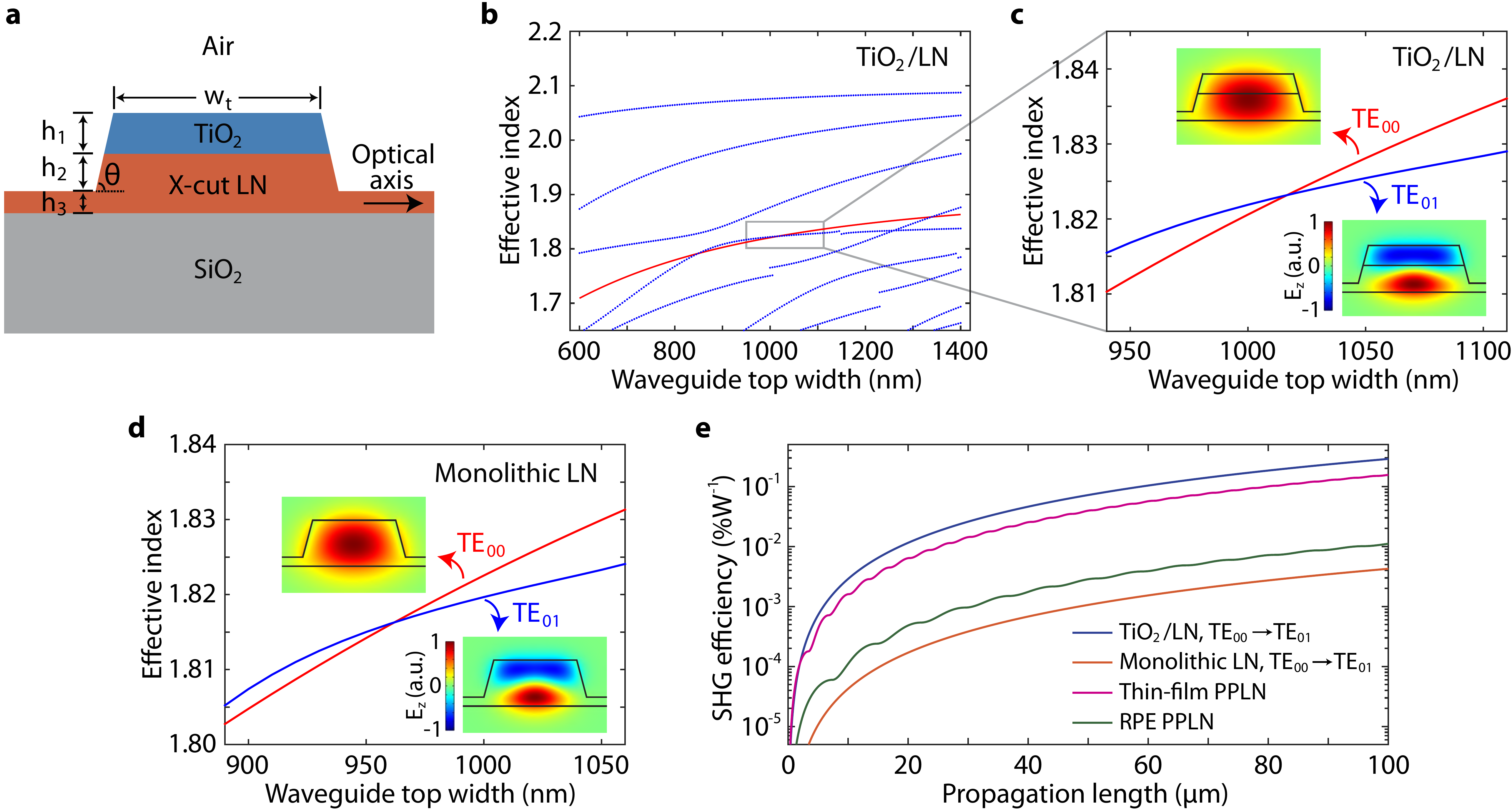}
	\caption{Design of a $\rm {TiO_2}$/LN semi-nonlinear waveguide for an enhanced SHG efficiency. {\bf a,} Schematic of the $\rm {TiO_2}$/LN waveguide. {\bf b,} Effective refractive indices of the $\rm {TiO_2}$/LN waveguide as functions of the waveguide top width $w_t$, simulated by the finite element method. Other waveguide dimensions are set as $h_1=220$ nm, $h_2=200$ nm, $h_3=100$ nm, and $\theta=75^\circ$. The red solid curve is for the TE$_{00}$ mode at the fundamental wavelength of 1550~nm, and the blue dotted curves are for different quasi-TE modes at the half wavelength of 775 nm. Discontinuities in some blue curves are due to coupling with certain quasi-TM modes (not shown). {\bf c,} Detailed effective indices of the $\rm {TiO_2}$/LN waveguide as functions of $w_t$, showing modal phase matching between TE$_{00}$ at 1550 nm and TE$_{01}$ at 775 nm around $w_t$=1020 nm. The insets show the optical field ($E_z$) profiles of both modes, with a shared color bar. {\bf d,} Detailed effective indices as functions of $w_t$, for a monolithic LN waveguide, showing phase matching between TE$_{00}$ at 1550 nm and TE$_{01}$ at 775 nm around $w_t=960$ nm. Other than a smaller phase-matched $w_t$ due to different material dispersions, the geometry parameters of the monolithic LN waveguide are the same as those of the $\rm {TiO_2}$/LN waveguide in {\bf b} and {\bf c}, while with the $\rm {TiO_2}$ layer replaced by LN. {\bf e,} Evolution of SHG efficiencies in a $\rm {TiO_2}$/LN semi-nonlinear waveguide (see {\bf c}), a monolithic LN waveguide (see {\bf d}), a thin-film PPLN waveguide (Ref.~\cite{Bowers16}), and a typical RPE PPLN waveguide (Ref.~\cite{Fejer02typical}), assuming no loss and no pump depletion for all types of waveguides, as well as ideally uniform poling for PPLN. } \label{Fig2}
\end{figure*}

Equation (\ref{eta}) shows that the efficiency of SHG depends essentially on the nonlinear susceptibility, the effective mode area, and the spatial mode overlap. In particular, Eq.~(\ref{zeta}) shows that the spatial mode overlap relies critically on the relative spatial symmetry between the FF and SH modes. Unfortunately, for a nonlinear waveguide, different-order guided modes generally exhibit very distinctive spatial symmetries, leading to a dramatically degraded spatial mode overlap. To illustrate this problem, we consider a nominal monolithic nanophotonic rib waveguide shown in Fig.~\ref{Fig1}a, where the core material exhibits a quadratic nonlinearity. An appropriate waveguide geometry will lead to exact phase matching between a fundamental TE$_{00}$ mode at a wavelength $\lambda$ and a high-order mode (TE$_{01}$ in our example) at the half wavelength $\lambda/2$. However, as shown in Fig.~\ref{Fig1}b,c, the electric field ($E_z$) of the SH mode changes its polarity across the waveguide core while that of the FF mode maintains a single polarity. As a result, the nonlinear parametric interaction in the upper half of the waveguide is out of phase with that in the lower half, and therefore they cancel with each other, leading to a very small net nonlinear effect. It is exactly this modal mismatch of spatial symmetries that undermines severely the nonlinear conversion efficiency in the majority of current nonlinear photonic chips \cite{Vuckovic09, Lipson11, Leo11, Tang11, Pavesi12, Noda14, Vuckovic14, Solomon14, Barclay16, Tang16Optica, Bres17, Pertsch15, Cheng16, Bowers16, Fathpour16PPLN, Loncar17OE, Luo17OE, Buse17, Luo18arXiv, Xiao18, Loncar18CLEO, Bowers18}, although they have shown the advantage of strong mode confinement, i.e. small mode areas. Consequently, a device has to rely on a long interaction length (or a high optical Q in the case of a resonator) to sustain the nonlinear process \cite{Tang16Optica}.

\begin{figure*}[t!]
	\centering\includegraphics[width=2.0\columnwidth]{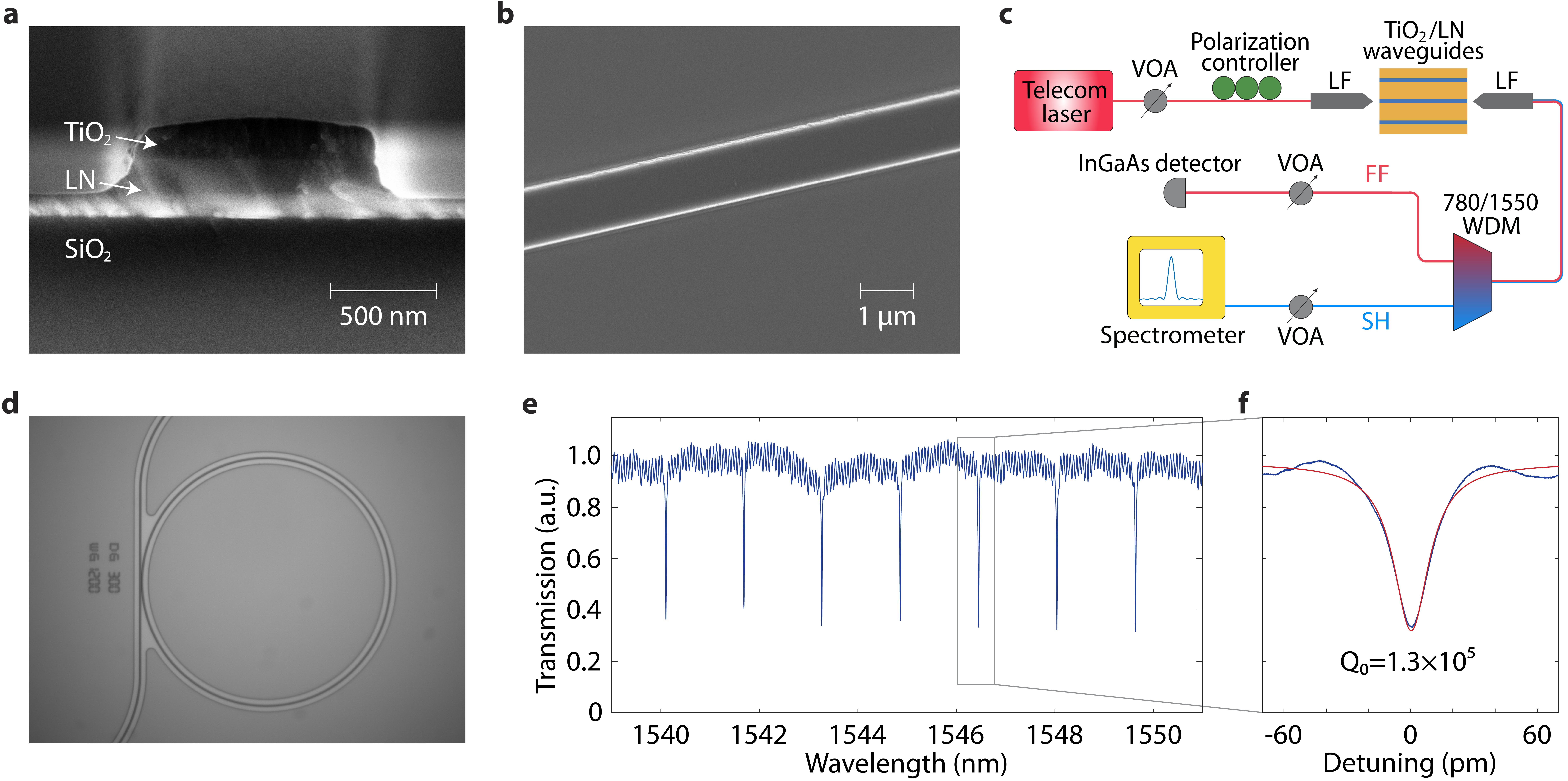}
	\caption{ Characterization of fabricated devices. {\bf a,} Scanning electron microscopy image of the facet of a fabricated $\rm {TiO_2}$/LN waveguide. {\bf b,} Top view of a section of the waveguide. {\bf c,} Schematic of the experimental setup for device characterization and SHG measurement. VOA: variable optical attenuator; LF: lensed fiber; WDM: wavelength division multiplexer. {\bf d,} Optical microscopy image of a $\rm {TiO_2}$/LN microring resonator with a radius of 100 $\mu$m. {\bf e,} Transmission spectrum of the microring resonator in the telecom band. {\bf f,} Detailed transmission spectrum of a typical resonance, with experimental data shown in blue and a fitting curve shown in red. } \label{Fig3}
\end{figure*}

To tackle this challenge, we propose semi-nonlinear waveguides which leverage the fact that modal phase matching depends only on the linear properties of the waveguide independent of its nonlinearity. Therefore, by breaking the spatial symmetry of the waveguide nonlinearity, we shall be able to significantly enhance the strength of the nonlinear interaction. Figure \ref{Fig1}d shows the schematic of a semi-nonlinear waveguide, which exhibits the same geometry as the monolithic waveguide shown in Fig.~\ref{Fig1}a, while the top part of the core is replaced by a linear material with the same refractive index, and a vanished $\chi^{(2)}$. Since its linear refractive index profile is the same as that of the monolithic waveguide, the semi-nonlinear waveguide will support an exactly same pair of phase-matched TE$_{00}$ mode at the FF and TE$_{01}$ mode at the SH, with mode profiles identical to those of the monolithic waveguide (see Fig.~\ref{Fig1}b,e). The thickness of the linear layer is chosen such that, for the SH mode TE$_{01}$ with two opposite polarities, $E_z \approx 0$ at the boundary between the two core materials, resulting in a single polarity inside each of the linear and nonlinear layers. As the top linear layer does not participate in the nonlinear interaction, the parametric process only has contribution from the bottom nonlinear layer, which will not be canceled, resulting in a remarkably enhanced net nonlinear effect [which manifests as a large value of $\zeta$ in Eq.~(\ref{zeta})]. The discussion above, for simplicity, has assumed an identical refractive index between the linear and nonlinear layers. In practice, the operation principle can be applied to two core materials with dissimilar refractive indices, since the semi-nonlinear waveguide offers plenty of degrees of freedom for engineering.

\section{Waveguide design}

To demonstrate the proposed principle, we design a semi-nonlinear waveguide that consists of a single-crystalline LN thin-film layer as the nonlinear medium and an amorphous $\rm {TiO_2}$ layer as the linear component. LN exhibits a large $\chi^{(2)}$ nonlinearity and a wide transparency window from ultraviolet to mid-infrared, and is an ideal medium for nonlinear parametric generation \cite{Fejer07, Shur15, Byer92, Watanabe93, Fejer95, Kato97, Fejer02typical, Fejer02higheff, Fejer10, Pertsch15, Cheng16, Bowers16, Fathpour16PPLN, Loncar17OE, Luo17OE, Buse17, Luo18arXiv, Xiao18, Loncar18CLEO}. $\rm {TiO_2}$ is chosen as the linear material for a few reasons. First, $\rm {TiO_2}$ deposited by physical vapor deposition is in an amorphous phase where the inversion symmetry leads to a vanished $\chi^{(2)}$. Second, our characterization shows that an amorphous $\rm {TiO_2}$ thin film has a relatively large refractive index of 2.137 at 1550 nm and 2.186 at 775 nm, making it suitable for strong optical confinement and flexible dispersion engineering. Third, $\rm {TiO_2}$ also has a large bandgap and has been demonstrated for high-quality waveguides at telecom and visible wavelengths \cite{Yoo13, Lipson13, Suntivich15}, enabling it to be a low-loss core material that guides both FF and SH light.

The designed $\rm {TiO_2}$/LN semi-nonlinear nanophotonic waveguide is schematically shown in Fig.~\ref{Fig2}a. X-cut LN is employed for its large second-order nonlinearity in the device plane, ideal for efficient type-0 processes involving quasi-TE modes. Numerical simulations show that the TE$_{00}$ mode at 1550~nm can be phase matched to certain quasi-TE modes at 775 nm with appropriate waveguide widths (see Fig.~\ref{Fig2}b). In particular, as shown in Fig.~\ref{Fig2}c, a waveguide width of $\sim$1020 nm is able to produce exact phase matching between the TE$_{00}$ mode at 1550 nm and the TE$_{01}$ at 775 nm, the latter of which exhibits an optical mode field (see Fig.~\ref{Fig2}c, insets) with two opposite polarities located separately in the linear and nonlinear core layers, a desired property we have shown in Fig.~\ref{Fig1}.

Detailed simulation shows that our waveguide exhibits a large overlap factor of $\zeta$=0.66, a value significantly beyond what is achievable in monolithic nanophotonic waveguides through modal phase matching \cite{Lipson11, Leo11, Tang11, Solomon14, Barclay16, Tang16Optica, Bres17, Pertsch15, Cheng16, Tang16Optica, Loncar17OE, Luo17OE, Buse17, Luo18arXiv, Xiao18, Bowers18}. Together with a small effective mode area of $A_{\rm eff} = 2.24~{\rm \mu m^2}$ and a significant quadratic nonlinearity of $d_{\rm eff} = d_{33}= 27~\rm pm/V$, our device is able to exhibit a theoretical normalized conversion efficiency as large as $\eta = 2900\%~\rm {\rm {W^{-1}} cm^{-2}}$. In comparison, a conventional monolithic LN waveguide with a similar geometry (see Fig.~\ref{Fig2}d) can only offer an efficiency of $\eta = 43\%~\rm {\rm {W^{-1}} cm^{-2}}$, directly showing the advantage of our proposed approach. Figure \ref{Fig2}e compares evolution of the SHG efficiency as the pump light propagates in several different waveguides. It shows clearly that the $\eta$ value in our semi-nonlinear waveguide is more than one order of magnitude larger than that of typical reverse-proton-exchanged (RPE) periodically-poled lithium niobate (PPLN) waveguides \cite{Fejer02typical} and that of monolithic LN nanophotonic waveguides \cite{Loncar17OE, Luo18arXiv}, and almost doubles that of PPLN thin films loaded with SiN ridges \cite{Bowers16}.

\begin{figure*}[t!]
	\centering\includegraphics[width=1.6\columnwidth]{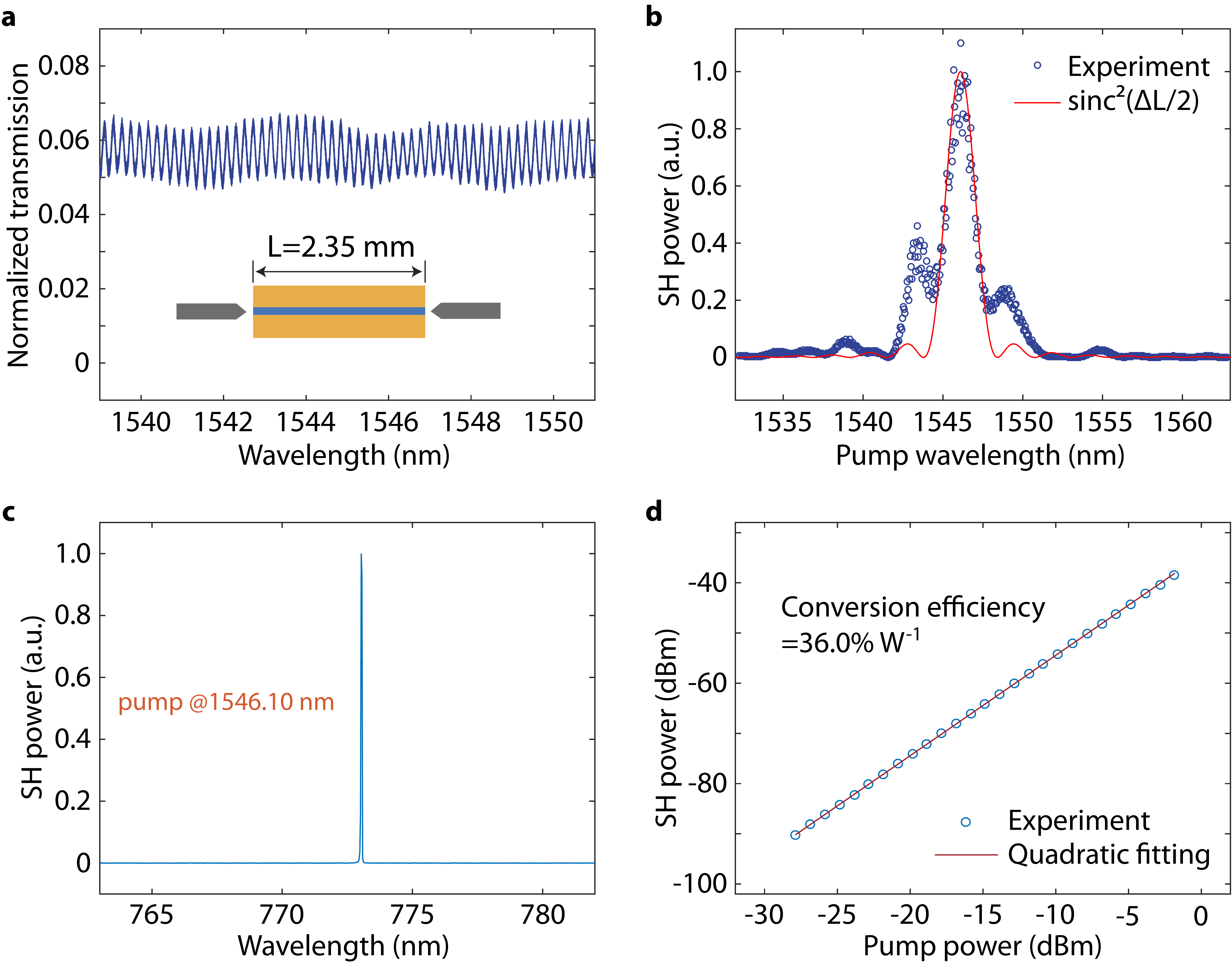}
	\caption{Measurement of SHG. {\bf a,} Transmission spectrum in the telecom band of a $\rm {TiO_2}$/LN straight waveguide with a length of $L$=2.35 mm. {\bf b,} Measured SHG spectrum (blue circles) of the $\rm {TiO_2}$/LN waveguide. The red curve shows ${\rm sinc}^2(\Delta L/2)=[\frac{\sin(\Delta L/2)}{\Delta L/2}]^2$ for comparison, where the phase mismatch $\Delta=\frac{4\pi}{\lambda}(n_1-n_2)$. {\bf c,} SHG spectrum of the device when it is pumped at a fixed wavelength of 1546.10 nm. {\bf d,} Power dependence of the SHG, with experimental data shown in blue and a theoretical quadratic fitting shown in red. The theoretical fitting indicates a conversion efficiency of 36.0$\%~\rm {W^{-1}} $.  } \label{Fig4}
\end{figure*}

\section{Linear optical properties}

To confirm our simulation results, we performed device fabrication, with the LN waveguide made through a standard top-down process \cite{Luo17OE, Luo18arXiv} and the $\rm {TiO_2}$ layer via a lift-off approach. Details of the fabrication procedure are discussed in Appendix. Figure \ref{Fig3}a shows the facet of a typical fabricated waveguide, where the $\rm {TiO_2}$ layer landed nicely on top of the etched LN rib waveguide, with a geometry close to our design (see Fig.~\ref{Fig2}c). Figure \ref{Fig3}b presents the top view of the waveguide, which shows a small sidewall roughness, implying a low propagation attenuation.

The device was tested with an experimental setup whose schematic is shown in Fig.~\ref{Fig3}c. To accurately characterize the waveguide propagation loss, we fabricated microring resonators, with an example shown in Fig.~\ref{Fig3}d. Figure~\ref{Fig3}e presents the laser-scanned transmission spectrum of a microring resonator with a radius of 100 ${\rm \mu m}$, which shows clearly the TE$_{00}$ mode family in the telecom band with a free-spectral range of 1.59~nm.
Figure~\ref{Fig3}f presents the detailed transmission spectrum of a typical resonance, which exhibits an intrinsic quality factor of 1.3$\times 10^5$, indicating a waveguide propagation loss of 3.2 dB/cm in the telecom band \cite{Dalton02}. For the SH mode around 775~nm, however, the current bus waveguide is very weakly coupled to the microring resonator, which makes it difficult to measure the quality factor in this waveband.

\section{Second-harmonic generation}

To demonstrate efficient SHG, we employed a straight waveguide with a length of about 2.35~mm, whose transmission in the telecom band is shown in Fig.~\ref{Fig4}a for the quasi-TE polarization. From Fig.~\ref{Fig4}a, we extract a fiber-to-fiber loss of 12.5~dB. Given a propagation loss of 3.2 dB/cm, we retrieve a fiber-to-chip coupling loss of 5.9 dB/facet. By scanning the pump laser wavelength, we were able to characterize the efficiency spectrum of SHG. One example is shown in Fig.~\ref{Fig4}b, which indicates a phase-matched pump wavelength of about 1546~nm. The main lobe of the recorded efficiency spectrum agrees well with the theoretical expectation from the sinc$^2$-function. The strong side lobes are likely introduced by slight non-uniformity of the waveguide (for example, potential thickness variations of the core layers).

By fixing the pump wavelength at 1546.10 nm where the peak conversion is located, we observed coherent radiation from its SH at 773.05 nm, shown as a sharp peak in Fig.~\ref{Fig4}c. The measured SH power shows a quadratic dependence on the pump power (see Fig.~\ref{Fig4}d), which agrees very well with the theoretical expectation. Fitting the experimental data, we obtained an on-chip conversion efficiency of 36.0\%~$\rm {W^{-1}}$ (see Fig.~\ref{Fig4}d), indicating an experimentally recorded normalized conversion efficiency of 650\%~$\rm {W^{-1}}cm^{-2}$. This value is about a 4-fold increase compared with the previous record of SHG in LN in the same waveband \cite{Fejer02higheff, Bowers16}.

The recorded SHG efficiency is smaller than the theoretical value given by $\eta L^2$ (=160\%~$\rm W^{-1}$), which is mainly due to the propagation losses of the waveguide. We have measured a propagation loss of 3.2 dB/cm for the FF mode. As a rough estimate, we assume that of the SH mode to be $\sim$12.8~dB/cm, since the propagation loss is dominated by Rayleigh scattering from the roughness of the waveguide surface, which scales with wavelength as $1/\lambda^2$ \cite{Tien71}. Consequently, the theoretical conversion efficiency is estimated to be $\sim$73.5\%~$\rm W^{-1}$ for the 2.35-mm-long waveguide, after we take into account the propagation losses. In addition, the slight waveguide non-uniformity could also impact the recorded conversion efficiency to a certain extent. As all these factors can be resolved with further optimization of the device fabrication (say, by using chemical-mechanical polishing to improve the surface smoothness and waveguide uniformity \cite{Buse17}), we expect that the measured conversion efficiency can be increased considerably in the near future.

\section{Conclusion and discussion}

In conclusion, we have proposed and demonstrated a universal design and operation principle for quadratic parametric processes on integrated photonic platforms, which is able to achieve exact phase matching with a large nonlinearity, a small mode area, and a large mode overlap factor for extremely efficient SHG. With this principle, we designed a $\rm {TiO_2}$/LN semi-nonlinear waveguide that is able to offer a theoretical conversion efficiency as high as 2900$\%~\rm {\rm {W^{-1}} cm^{-2}}$, more than one order of magnitude higher than those achievable in RPE PPLN and monolithic on-chip LN waveguides \cite{Fejer02typical, Loncar17OE, Luo18arXiv}. The highly efficient parametric generation enabled us to record a SHG efficiency of 36\%~${\rm W^{-1}}$ inside a waveguide only 2.35~mm long, corresponding to a normalized efficiency of 650\%~${\rm W^{-1} cm^{-2}}$.

The SHG efficiency of our device is comparable with those of recently reported on-chip PPLN waveguides \cite{Bowers16, Loncar18CLEO}. However, our demonstrated approach is free from the complicated periodic poling process. Domain engineering has been regarded as the holy grail for quadratic nonlinear photonics \cite{Fejer07, Helmy11, Shur15}, which, however, is material selective. Our proposed approach, instead, can be universally applied to any on-chip quadratic nonlinear platforms, including for example, certain dielectrics \cite{Lipson11, Bres17}, group IV \cite{Pavesi12, Noda14}, III-V \cite{Tang11, Tang16Optica, Bowers18}, and II-VI \cite{Helmy11} semiconductor chips to which domain engineering is challenging to apply. Therefore, our proposed approach may open up a great avenue towards extreme nonlinear and quantum optics with ultra-high nonlinear conversion efficiencies that are promising for broad applications in energy efficient nonlinear and quantum photonic signal processing.

On the other hand, as shown in Fig.~\ref{Fig1}, the proposed semi-nonlinear nanophotonic waveguide intriguingly separates the waveguide into a nonlinear part that experiences nonlinear conversion gain and a linear part that experiences only linear propagation losses, which forms a natural parity-time-symmetric system that is of great potential for non-Hermitian photonic applications \cite{Feng17, Christodoulides18}.

\section*{Appendix}
\textbf{Device Fabrication.} We started from a LN-on-insulator wafer by NANOLN, with 300 nm of X-cut LN sitting on 2-$\rm \mu m$-thick buried oxide and a silicon substrate. LN waveguides were patterned by electron-beam lithography (EBL), with ZEP520A as the resist. After etching LN waveguides with ion milling, we removed residual resist with oxygen plasma followed by diluted hydrofluoric acid. Then, in order to pattern the $\rm {TiO_2}$ layer on top of the etched LN waveguides, we employed aligned EBL, after which we used physical vapor deposition to deposit $\rm {TiO_2}$. Next, we soaked the wafer in 1165 resist remover, which resolved the resist, leaving us the heterogeneous $\rm {TiO_2}$/LN waveguides. Finally, the device chip was hand-cleaved for light coupling into the waveguides.

\textbf{Experimental Setup.} Pump light from a continuous-wave tunable telecom-band laser was coupled into the device chip through a lensed fiber. At the waveguide output, FF pump light was collected together with the SH light by a second lensed fiber. Then, light at the two wavebands were separated by a 780/1550 WDM, after which the FF light was directed to an InGaAs detector for characterization, while the SH light was sent to a spectrometer for detection. A fiber polarization controller was used to achieve optimal coupling of the wanted polarization mode, and VOAs were employed to study the power dependence of SHG.


\section*{Acknowledgments}
The authors thank Chengyu Liu at Cornell University for helpful discussions on fabrication of $\rm TiO_2$. This work was supported in part by the National Science Foundation under Grant No.~ECCS-1641099, ECCS-1509749 and ECCS-1810169, by the Defense Threat Reduction Agency under the Grant No. HDTRA1827912, and by the Defense Advanced Research Projects Agency SCOUT program through grant number W31P4Q-15-1-0007 from the U.S. Army Aviation and Missile Research, Development, and Engineering Center (AMRDEC). The project or effort depicted was or is sponsored by the Department of the Defense, Defense Threat Reduction Agency. The content of the information does not necessarily reflect the position or the policy of the federal government, and no official endorsement should be inferred. This work was performed in part at the Cornell NanoScale Facility, a member of the National Nanotechnology Coordinated Infrastructure (National Science Foundation, ECCS-1542081), and at the Cornell Center for Materials Research (National Science Foundation, DMR-1719875).







\begin{thebibliography}{99}
		
\bibitem{BoydBook}
	Boyd, R. W. \textit{Nonlinear Optics (3rd edn.).} (Academic, NewYork, 2008).
\bibitem{Bloembergen00}
 	Bloembergen N. Nonlinear optics: past, present, and future. \textit{IEEE J. Sel. Top. Quantum Electron.} {\bf 6,} 876-880 (2000).
\bibitem{Fejer06}
	Carsten, L., Kumar, S., McGeehan, J. E., Willner, A. E. \& Fejer, M. M. All-optical signal processing
	using $\chi^{(2)}$ nonlinearities in guided-wave devices. \textit{J. Lightwave Technol.} {\bf 24,} 2579-2592 (2006).	
\bibitem{Willner14}
	Willner, A. E., Khaleghi, S., Chitgarha, M. R. \& Yilmaz, O. F. All-optical signal processing. \textit{J. Lightwave Technol.} {\bf 32,} 660-680 (2014).
\bibitem{Dunn99}
    Dunn, M. H. \& Ebrahimzadeh, M. Parametric generation of tunable light from continuous-wave to femtosecond pulses. \textit{Science} {\bf 286,} 1513-1517 (1999).
\bibitem{Cundiff03}
	Cundiff, S. T. \& Ye, J. Colloquium: Femtosecond optical frequency combs. \textit{Rev. Mod. Phys.} {\bf 75,} 325-342 (2003).
\bibitem{Dong11}
	Campagnola, P. J. \& Dong, C. Y. Second harmonic generation microscopy: principles and applications to disease diagnosis, \textit{Laser Photonics Rev.} {\bf 5,} 13-26 (2011).
\bibitem{Pan12}
	Pan, J.-W. \textit{et al.} Multiphoton entanglement and interferometry. \textit{Rev. Mod. Phys.} {\bf 84,} 777-838 (2012).
\bibitem{Lipson11}
	Levy, J. S., Foster, M. A., Gaeta, A. L. \& Lipson, M. Harmonic generation in silicon nitride ring resonators. \textit{Opt. Express} {\bf 19,} 11415-11421 (2011).
\bibitem{Bres17}
	Billat, A. \textit{et al.} Large second harmonic generation enhancement in Si$_3$N$_4$ waveguides by all-optically induced quasi-phase-matching, \textit{Nature Commun.} {\bf 8,} 1016 (2017).	
\bibitem{Pavesi12}
	Cazzanelli, M. \textit{et al.} Second-harmonic generation in silicon waveguides strained by silicon nitride, \textit{Nature Mater.} {\bf 11,} 148-154 (2012).
\bibitem{Noda14}
	Yamada, S. \textit{et al.} Second-harmonic generation in a silicon-carbide-based photonic crystal nanocavity. \textit{Opt. Lett.} {\bf 39,} 1768-1771 (2014).
\bibitem{Barclay16}
	Lake, D. P. \textit{et al.} Efficient telecom to visible wavelength conversion in doubly resonant gallium phosphide microdisks, \textit{Appl. Phys. Lett.} {\bf 108,} 031109 (2016).
\bibitem{Vuckovic09}
	Rivoire, K., Lin, Z., Hatami, F., Masselink, W. T. \& Vu{\v{c}}kovi{\'c}, J. Second harmonic generation in gallium phosphide photonic crystal nanocavities with ultralow continuous wave pump power, \textit{Opt. Express} {\bf 17,} 22609-22615 (2009).
\bibitem{Vuckovic14}
	Buckley, S. \textit{et al.} Nonlinear frequency conversion using high-quality modes in GaAs nanobeam cavities, \textit{Opt. Lett.} {\bf 39,} 5673-5676 (2014).	
\bibitem{Leo11}
	Savanier, M. \textit{et al.}  Large second-harmonic generation at 1.55 ${\rm \mu}$m in oxidized AlGaAs waveguides, \textit{Opt. Lett.} {\bf 36,} 2955-2957 (2011).
\bibitem{Tang11}
	Xiong C. \textit{et al.}  Integrated GaN photonic circuits on silicon (100) for second harmonic generation, \textit{Opt. Express} {\bf 19,} 10462-10470 (2011).
\bibitem{Solomon14}
	Kuo, P. S., Bravo-Abad, J. \& Solomon, G. S. Second-harmonic generation using $\bar{4}$-quasi-phase matching in a GaAs whispersing-gallery-mode microcavity, \textit{Nature Commun.} {\bf 5,} 3109 (2014).
\bibitem{Tang16Optica}
	Guo, X., Zou, C.-L. \& Tang, H. X. Second-harmonic generation in aluminum nitride microrings with 2500\%/W conversion efficiency, \textit{Optica} {\bf 3,} 1126-1131 (2016).
\bibitem{Bowers18}
	Chang, L. \textit{et al.} Heterogeneously integrated GaAs waveguides on insulator for efficient frequency conversion. Preprint at http://arXiv:1805.09379 (2018).
\bibitem{Pertsch15}
	Geiss, R. \textit{et al.} Fabrication of nanoscale lithium niobate waveguides for second-harmonic generation. \textit{Opt. Lett.} {\bf 40,} 2715-2718 (2015).	
\bibitem{Cheng16}
	Lin, J. \textit{et al.} Phase-matched second-harmonic generation in an on-chip LiNbO$_3$ microresonator. \textit{Phys. Rev. Appl.} {\bf 6,} 014002 (2016).	
\bibitem{Loncar17OE}
	Wang, C. \textit{et al.} Second harmonic generation in nanostructured thin-film lithium niobate waveguides. \textit{Opt. Express} {\bf 25,} 6963-6973 (2017).
\bibitem{Luo17OE}
	Luo, R. \textit{et al.} On-chip second-harmonic generation and broadband parametric down-conversion in a lithium niobate microresonator. \textit{Opt. Express} {\bf 25,} 24531-24539 (2017).		
\bibitem{Buse17}
	Wolf, R., Breunig, I., Zappe, H. \& Buse, K. Cascaded second-order optical nonlinearities in on-chip micro rings. \textit{Opt. Express} {\bf 25,} 29927-29933 (2017).
\bibitem{Luo18arXiv}
	Luo, R., He, Y., Liang, H., Li, M. \& Lin, Q. highly tunable efficient second-harmonic generation in a lithium niobate nanophotonic waveguide. Preprint at http://arXiv:1804.03621 (2018).	
\bibitem{Xiao18}
	Wang, L. \textit{et al.} High-Q chaotic lithium niobate microdisk cavity. \textit{Opt. Lett.} {\bf 43,} 2917-2920 (2018).
\bibitem{Bowers16}
	Lin, C. \textit{et al.} Thin film wavelength converters for photonic integrated circuits. \textit{Optica} {\bf 3,} 531-535 (2016).
\bibitem{Fathpour16PPLN}
	Rao, A. \textit{et al.} Second-harmonic generation in periodically-poled thin film lithium niobate wafer-bonded on silicon. \textit{Opt. Express} {\bf 24,} 29941-29947 (2016).
\bibitem{Loncar18CLEO}
	Wang, C. \textit{et al.} Second-harmonic generation in nanophotonic PPLN waveguides with ultrahigh efficiencies. in \textit{Conference on Lasers and Electro-Optics} (Optical Society of America, 2018) p. JTh5A.2.
\bibitem{Helmy11}
	Helmy, A. S. \textit{et al.} Recent advances in phase matching of second-order nonlinearities in monolithic semiconductor waveguides. \textit{Laser Photonics Rev.} {\bf 5,} 272-286 (2011).
\bibitem{Fejer07}
	Hum, D. S. \& Fejer, M. M. Quasi-phasematching. \textit{C. R. Physique} {\bf 8,} 180-198 (2007).
\bibitem{Shur15}
	Shur, V. Y., Akhmatkhanov, A. R. \& Baturin, I. S. Microand nano-domain engineering in lithium niobate. \textit{Appl. Phys. Rev.} {\bf 2,} 040604 (2015).
\bibitem{Byer92}
	Fejer, M. M., Magel, G. A., Jundt, D. H. \& Byer, R. L. Quasi-phase-matched second harmonic generation: tuning and tolerances. \textit{IEEE J. Quantum Electron.} {\bf 28,} 2631-2654 (1992).
\bibitem{Watanabe93}
	Yamada, M., Nada, N., Saitoh, M., \& Watanabe, K. First-order quasi-phase matched LiNbO$_3$ waveguide periodically poled by applying an external field for efficient blue second-harmonic generation. \textit{Appl. Phys. Lett.} {\bf 62,} 435-436 (1993).	
\bibitem{Fejer95}
	Myers, L. E. \textit{et al.} Quasi-phase-matched optical parametric oscillators in bulk periodically poled LiNbO$_3$. \textit{J. Opt. Soc. Am. B} {\bf 12,} 2102-2116 (1995).	
\bibitem{Kato97}
	Mizuuchi, K., Ohta, H., Yamamoto, K. \& Kato, M. Second-harmonic generation with a high-index-clad waveguide. \textit{Opt. Lett.} {\bf 22,} 1217-1219 (1997).	
\bibitem{Fejer02typical}
	Parameswaran, K. R., Kurz, J. R., Roussev, R. V. \& Fejer, M. M. Observation of 99\% pump depletion in single-pass second-harmonic generation in a periodically poled lithium niobate waveguide. \textit{Opt. Lett.} {\bf 27,} 43-45 (2002).
\bibitem{Fejer02higheff}
	Parameswaran, K. R. \textit{et al.} Highly efficient second-harmonic generation in buried waveguides formed by annealed and reverse proton exchange in periodically poled lithium niobate. \textit{Opt. Lett.} {\bf 27,} 179-181 (2002).
\bibitem{Fejer10}
	Pelc, J. S., Langrock, C., Zhang, Q. \& Fejer, M. M. Influence of domain disorder on parametric noise in quasi-phase-matched quantum frequency converters \textit{Opt. Lett.} {\bf 35,} 2804-2806 (2010).
\bibitem{Yoo95}
	Yoo, S. J. B., Bhat, R., Caneau, C. \& Koza, M. A. Quasi-phase-matched second-harmonic generation in AlGaAs waveguides with periodic domain inversion achieved by
	wafer-bonding. \textit{Appl. Phys. Lett.} {\bf 66,} 3410-3412 (1995).
\bibitem{Fejer01}
	Eyres, L. A. \textit{et al.} All-epitaxial fabrication of thick, orientation-patterned GaAs films for nonlinear optical frequency conversion. \textit{Appl. Phys. Lett.} {\bf 79,} 904-906 (2001).
\bibitem{Tassev13}
	Tassev, V. \textit{et al.} Progress in orientation-patterned GaP for next-generation nonlinear optical devices. \textit{Proc. SPIE} {\bf 8604}, 86040V (2013).
\bibitem{Yoo13}
	Djordjevic, S. S. \textit{et al.} CMOS-compatible, athermal silicon ring modulators clad with titanium dioxide \textit{Opt. Express} {\bf 21,} 13958-13968 (2013).
\bibitem{Lipson13}
	Guha, B., Cardenas, J. \& Lipson, M. Athermal silicon microring resonators with titanium oxide cladding. \textit{Opt. Express} {\bf 21,} 26557-26563 (2013).
\bibitem{Suntivich15}	
	Evans, C. E., Liu, C. \& Suntivich, J. Low-loss titanium dioxide waveguides and resonators using a dielectric lift-off fabrication process. \textit{Opt. Express} {\bf 23,} 11160-11169 (2015).
\bibitem{Dalton02}
	Rabiei, P., Steier, W. H., Zhang, C. \& Dalton, L. R. Polymer micro-ring filters and modulators. \textit{J. Light. Technol.} {\bf 20,} 1968-1975 (2002).
\bibitem{Tien71}
	Tien, P. K. Light waves in thin films and integrated optics. \textit{Appl. Opt.} {\bf 10,} 2395-2413 (1971).	
\bibitem{Feng17}
	Feng, L., El-Ganainy, R. \& Ge, L. Non-Hermitian photonics based on parity-time symmetry. \textit{Nature Photon.} {\bf 11,} 752-762 (2017).
\bibitem{Christodoulides18}
	El-Ganainy, R. \textit{et al.} Non-Hermitian physics and PT symmetry. \textit{Nature Phys.} {\bf 14,} 11-19 (2018).	
	
\end{thebibliography}
\end{document}